\documentclass[a4paper,superscriptaddress,twocolumn,prb]{revtex4}
\usepackage{amsmath}
\usepackage{graphicx}
\usepackage[english]{babel}

\usepackage[FIGTOPCAP,raggedright,normalsize,tight]{subfigure}

\makeindex             

\begin{document}

\title{Graphene edge structures: Folding, scrolling, tubing, rippling  and twisting}

\author{Viktoria V. Ivanovskaya}
\email{v.ivanovskaya@gmail.com}
\author{Philipp Wagner}
\affiliation{Institut des Mat\'eriaux Jean Rouxel (IMN), CNRS UMR 6502, Universit\'e de Nantes, 44322 Nantes, France}

\author{Alberto Zobelli}
\affiliation{Laboratoire de Physique des Solides, Univ. Paris-Sud, CNRS UMR 8502, 91405, Orsay, France}

\author{Irene Suarez-Martinez}
\affiliation{Nanochemistry Research Institute, Curtin University of Technology, Perth, Western Australia 6845, Australia}

\author{Abu Yaya}
\author{Christopher P. Ewels}
\email{chris.ewels@cnrs-imn.fr} 
\affiliation{Institut des Mat\'eriaux Jean Rouxel (IMN), CNRS UMR 6502, Universit\'e de Nantes, 44322 Nantes, France}

\begin{abstract}
Conventional three-dimensional crystal lattices are terminated by surfaces, which can demonstrate complex rebonding and rehybridisation, localised strain and dislocation formation.  Two dimensional crystal lattices, of which graphene is the archetype, are terminated by lines.  The additional available dimension at such interfaces opens up a range of new topological interface possibilities.  We show that graphene sheet edges can adopt a range of topological distortions depending on their nature.  Rehybridisation, local bond reordering, chemical functionalisation with bulky, charged, or multi-functional groups can lead to edge buckling to relieve strain, folding, rolling and even tube formation.  We discuss the topological possibilities at a 2D graphene edge, and under what circumstances we expect different edge topologies to occur.  Density functional calculations are used to explore in more depth different graphene edge types.
\end{abstract}

\maketitle

\section{Introduction}
A finite material is necessarily terminated by an interface.  While the bulk of a crystalline material respects crystal symmetry, this symmetry is broken at the interface.  Material interfaces are thus heterogenous and generally more reactive than the material bulk.  Symmetry breaking can lead to imbalanced local strain which needs compensating in some way, typically through interface relaxation but also potentially through dislocation creation.  Interfaces also create dangling bonds and enhanced reactivity, which once again can be mitigated through various effects such as chemical rehybridisation, interface reconstruction, and chemical functionalisation.

In the case of three-dimensional bulk solids their interfaces are two dimensional {\it surfaces}, but for two-dimensional materials such as graphene, their interfaces are one-dimensional {\it lines}.  As we discuss in this article, this difference in dimensionality means that graphene interfaces have a number of possible topological distortions and relaxation modes which are not available at `classic' surfaces of three-dimensional materials.  The result is a rich variety of potential interface types in graphene, all of which can radically alter the properties of the material, even at quite long range from the interface.

In the following article we discuss different interface behaviour in graphene.  We start by discussing more classic interface behaviours, giving examples of rehybridisation, interface reconstruction, restructuring and chemical functionalisation.  We then consider key new topological distortions that exploit the additional available third dimension.  For this it is useful to consider a geometrically anisotropic sample of graphene such as a graphene nanoribbon.  We take here the example of a graphene ribbon of infinite length, {\it i.e.} with two principle axes, one along the ribbon length and one orthogonal to it, both of which can demonstrate distinct topological edge distortions.

\section{Topological distortions within the graphene plane: Flat edges}
By flat edges we imply restricting the edge to remain within the graphene plane. In this case edge effects are direct one-dimensional analogues of surface behaviour in three-dimensional crystals.

\begin{figure*}

\includegraphics[scale=.25]{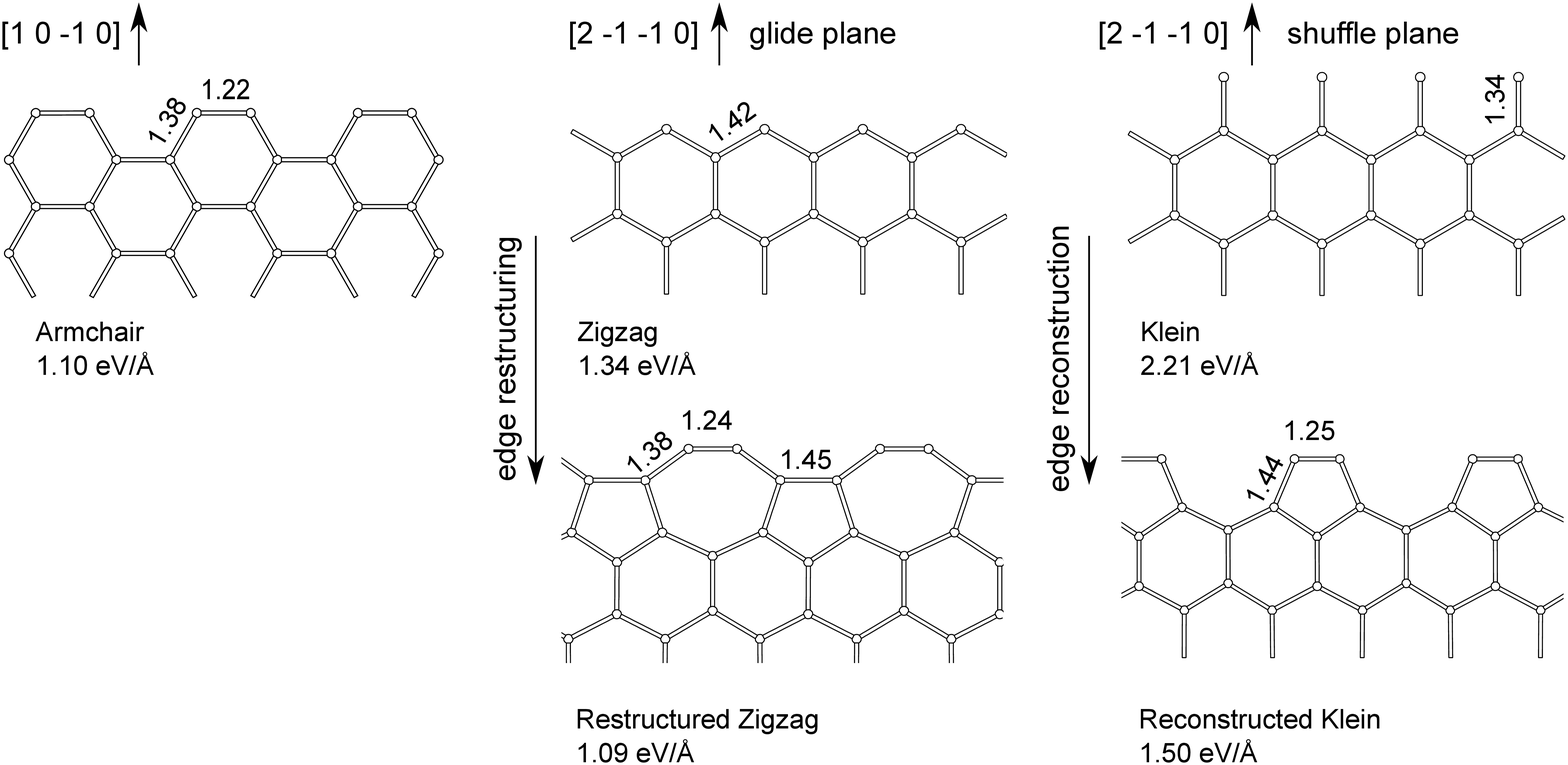}
\caption{Different edge structures for flat unterminated graphene edge (adapted from Supplementary Materials, Reference \citet{ourPRL}). (left) Armchair edge showing rehybridisation, (center) Zig-zag edge showing restructuring, with alternate edge bonds rotated about their bond centers, and (right) Klein edge showing edge reconstruction through pairwise rebonding. Edge formation energies (eV/\AA) are
indicated below each structure.}
\label{flatedges}    
\end{figure*}

\subsection{Rehybridisation}
Cutting carbon along the armchair, or `boat', direction, $[1\,0\, \overline{1}\, 0]$ results in a series of atom pairs.  These are able to rehybridise from sp$^2$ in the graphene bulk, towards sp triple bonding to stabilise the edge.  This shortens the bond from 1.412~\AA~ in the graphene bulk to 1.22~\AA, creating one of the most stable non-functionalised graphene edges, with a formation energy of only 1.10~eV/\AA \cite{ourPRL}, as shown in Figure~\ref{flatedges}(left).

\subsection{Interface reconstruction}
Dangling bonds on a two dimensional surface can be saturated through local rebonding between atoms at the interface.  This results in a surface superlattice containing more than one bulk lattice vector in at least one direction; for example the $(1\,0\, 0)$ surface of Si consists of Si atoms with three neighbours which stabilise by bonding in pairs, creating a $[2 \times 1]$ surface reconstruction.

A direct analogy at a graphene edge can be observed for the so-called Klein edge \cite{Klein}, cutting graphene along the $[2\, \overline{1}\, \overline{1}\, 0]$ direction in the shuffle plane. The result is a line of chemically unstable singly coordinated carbon atoms with correspondingly high edge formation energy (2.21 eV/\AA \cite{ourPRL}). Allowing symmetry breaking along the edge direction results in pairwise reconstruction, giving the graphene equivalent of a $[2 \times 1]$ surface reconstruction (a $[2]$ edge reconstruction) which drops the edge formation energy to 1.50 eV/\AA \cite{ourPRL}, as shown in Figure~\ref{flatedges}(right).

\subsection{Interface restructuring}
In some ways a subset of interface reconstruction, this class of edges undergo extensive reordering in the interface layer in order to create a more energetically favourable surface structure.  Cutting graphene once again along the $[2\, \overline{1}\, \overline{1}\, 0]$ direction, but this time in the glide plane, results in a zig-zag terminated edge, slightly more stable than the reconstructed Klein discussed above (1.34eV/\AA \cite{ourPRL}).  This edge is metallic and cannot stabilise through local rehybridisation.  However a series of 90$^\circ$ bond rotations along the edge around the bond centres for alternate bonds changes the crystal lattice from a hexagonal array to a periodic array of alternating pentagons and hexagons \cite{Pekka}, a linear analogue of the proposed layered Haeckelite structures.  In this case the edge atoms are now arranged pairwise, similarly to the armchair edges discussed above, and the edge carbon atoms can thus once again rehybridise, resulting in a very stable edge structure with formation energy of only 1.09eV/\AA \cite{ourPRL}, as shown in Figure~\ref{flatedges}(center).

\section{Topological distortions orthogonal to the ribbon axis: Folding, scrolling and tubing}

We now consider possible edge distortions which have no direct analogy in surface termination of three dimensional materials, {\it i.e.} these interface modes are unique to two dimensional layered materials.

\begin{figure}[!]

\begin{center}
\includegraphics[scale=1.30]{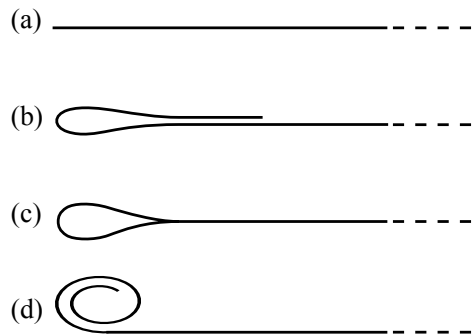}
\end{center}

\caption{Schematic cross-section showing four possible out-of-plane distortions of graphene sheet edges orthogonal to the edge line: (a) flat, (b) folded, (c) folded and rebonded (tubing) and (d) scrolled edges.}
\label{edgestypes}    
\end{figure}

If we consider the graphene nanoribbon in cross-section looking along the axis, there are four distinct classes of edge distortion possible.  The first is to simply remain planar, as discussed above (Figure~\ref{edgestypes}a).  However if we consider a thought experiment where the edge is now pulled up out of the graphene plane and back above the graphene layer, there are three possible structures depending on the angle at which the edge then approaches the graphene surface below it.  If the approach angle is less than 90$^\circ$ the edge folds back on itself, resulting in a bilayer structure (Figure~\ref{edgestypes}b).  If the approach angle is 90$^\circ$ the edge will bond into the sheet below (Figure~\ref{edgestypes}c), creating a pseudo-nanotube at the graphene edge, and if the approach angle is greater than 90$^\circ$ the edge folds back in on itself, creating a scroll (Figure~\ref{edgestypes}d).  We now discuss each of these in more detail.

\subsection{Folding}
The flexibility of graphene layers promotes the formation of self-folded
nanostructures (Figure~\ref{edgestypes}b). High resolution transmission electron microscopy (HRTEM) studies
\cite{PhysRevLett-Kazu, Warner-2, Warner, Huang,girit2009graphene,gass2008free,
meyer2007structure, meyer2007roughness}
have shown that graphene edges can fold back on themselves.
The energetic cost of bending the layer is compensated by Van der
Waals interactions in the stacked region\cite{rotkin2002analysis}.
Multiple folding of graphene results in multi-layer regions and highly
curved folding edges\cite{grafold}. The critical self-folded length at which point folding
becomes thermodynamically favourable has been theoretically deduced\cite{cranford2009meso}.

Graphene folding can occur in any direction, however the interaction
between the folded layers depends on the resultant atomic stacking. Scanning tunneling
microscopy (STM) and HRTEM studies suggest the possibility of both AA and AB
stacking for folded graphene layers \cite{PhysRevLett-Kazu,roy1998study}.
Energetic comparison for folded structures demonstrates the preference of
graphene folding: the global minimum is associated with AB stacking of the
entire flat region, while the local minimum with a mixture of AB and
incommensurate stacking occurs in the presence of a small twist of folded
graphene\cite{feng2009geometric,zhang2010free}.

Folded graphene edges present a combination of a nanotube-like and
multilayer graphene structures which give rise to peculiar electronic properties\cite{feng2009geometric,grafold}.
Nonetheless such edges are potentially very stable and commonly seen in multi-layer graphitic stacked materials.
In such cases it is also common to observe more than one sheet folded simultaneously together resulting in a bi-
or multi-layer fold.

\subsection{Scrolling}
Folding and consequential rolling up of a graphene layer into a spiral structure
at the sheet edge leads to the formation of a scroll (Figure~\ref{edgestypes}d).
A large variety of possible scroll structures can be obtained by coiling a
single or
multiple graphene sheets, changing the number of coils and
sliding relatively adjacent layers. The open and highly modulable structure
of scrolls suggests potential applications for hydrogen storage or for use
as ion
channels. \cite{mpourmpakis2007carbon} Experimentally, scrolls have been
obtained via arc discharge, chemical treatment of
graphite {} or graphene {} but an easy and reproducible route for scroll
synthesis still remains to be developed.

Several theoretical works have investigated the formation and structural stability of graphene
scrolls\cite{suarez2007dislocations, xu2010geometry}.
Similarly to graphene folding, the formation of scrolls is dominated by
two major energy contributions: the
energy increase due to the bending of the planar layer
and the energy gain due to the van der Waals interactions
between the rolled layers. Beyond a critical diameter value
these scrolled structures can be energetically more stable than the equivalent
planar configurations\cite{martins2010curved,braga2004structure}, however in
order to obtain a scroll a large energy barrier due to the bending rigidity
should be overcome. 

Scroll formation occurs spontaneously when a critical overlap between layers is achieved for the
partially curled sheet, and interestingly scroll unwinding has been observed during
charge injection \cite{braga2004structure}.  Thus, electrostatic control of the
wrapping appears feasible, opening the way to possible technological
applications\cite{fogler2010effect}. 

The minimum innermost radius of nanoscrolls was also calculated
\cite{chen2007structural,braga2004structure,pan2005ab}, and is experimentally observed
to be in the range 20-50\AA\cite{chen2007structural}.
Thus scrolling will be limited at graphene edges of larger flakes and will not be
important for graphene nanoribbons.

Similarly to nanotubes, the electronic structure of nanoscrolls has been shown to depend on their chirality ($n$,$m$)\cite{chen2007structural}. Armchair
nanoscrolls were found to be metallic or semimetallic depending on their sizes.
Metallic scrolls have a larger density of states at the Fermi level than
metallic single-walled nanotubes. Zigzag nanoscrolls were found to be
semiconducting with energy gap much smaller than corresponding zigzag nanotubes.
The optical properties of carbon nanoscrolls have been studied as well: the
calculated reflection spectra and loss function showed features of both
single-wall carbon and multiwall carbon nanotubes \cite{pan2005ab}.

\subsection{Tubing}
By folding an unterminated graphene edge back on itself and rebonding it into the graphene layer it is possible to create
`nanotube-terminated' edges\cite{ourPRL} (Figure~\ref{edgestypes}c). 
Depending on the edge chirality, zigzag and armchair edges lead to the formation of armchair and
zigzag tube-terminated edges, respectively. 
The edge dangling bonds are thus replaced with sp$^3$-hybridised
carbon atoms. For sufficiently large tubes (armchair tubes larger than (8,8) and
zigzag larger than (14,0)) this results in lower formation energies than any other
non-functionalised edge structure discussed here.

Simulated high resolution trasmission electron microscopy images are found to be
similar to those of free standing edges, the primary difference being minor
variations in image contrast, suggesting `tubed' edges may be difficult to distinguish experimentally.

\begin{figure}[!]

\begin{center}
\includegraphics[scale=.35]{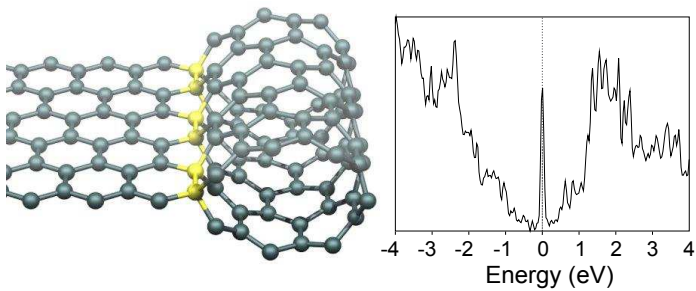}
\end{center}

\caption{`Tubed' graphene edge, where a non-functionalised edge is rolled back on itself and bonds into the layer below, creating a line of sp$^3$-hybridised carbon atoms (marked in yellow), in this case for a (8,0) zig-zag `tube'. (right) Density of states for a (8,8) armchair 'tube' edge, where the strong Fermi level peak corresponding to metallic states along the carbon atoms neighbouring the sp$^3$-bonded atoms is clearly visible.  Figure adapted from Reference \citet{ourPRL}.  }
\label{tube-edge}    
\end{figure}

The electronic properties of such tube terminated edges present an
interesting combination of graphene and nanotube behaviour.
Rolled zigzag edges serve as metallic conduction channels,
separated from the neighbouring bulk graphene by a chain of insulating
sp$^3$-carbon atoms, and introduce Van Hove singularities into the graphene
density of states (Figure \ref{tube-edge}). They may provide a way to stabilise and protect from chemical
attack the disperse Fermi level state seen along metallic zigzag
edges.\cite{ourPRL}

\section{Topological distortions parallel to the ribbon axis: Rippling and twisting}

\subsection{Rippling via functionalisation}
\label{ripples}

Pristine graphene edges have dangling bonds at the edge atoms. Simple H-termination is the simplest way to saturate these dangling bonds \cite{Wassmann2008}. With this approach very little strain is induced to the edge. 
Adding different atoms to the edge such as -F, -Cl or more complex functional groups such as -OH or -SH changes this simple picture. 
In general most functional groups add significant strain along the ribbon edge, through steric hindrance, electrostatic repulsion between groups, inter-group bonding (such as hydrogen bonding), etc. 

This strain is energetically unfavourable, and can be relieved via out-of-plane distortion\cite{Wagner2011}. 
Specifically, hydroxyl (-OH) terminated graphene nanoribbons of different widths (notably armchair edges) have been shown to compensate the induced strain by forming a localised static ripple along the edge \cite{Wagner2011}. This rippled edge is more stable than any flat configuration. The strain is relieved via distorting the edge carbon hexagons pairwise up and down periodically. The ripple is localised at the edge, confined by the sp$^2$ $\pi$ - system of graphene trying to stay flat.  These ripples are of a different length scale and underlying mechanism to previously observed statistical thermal fluctuations, and bending of the graphene surface via the physisorption of molecules such as H$_2$O \cite{Fasolino2007,Thompson-Flagg2009}.

\begin{figure}[!]

\subfigure [][] {\includegraphics[scale=.15]{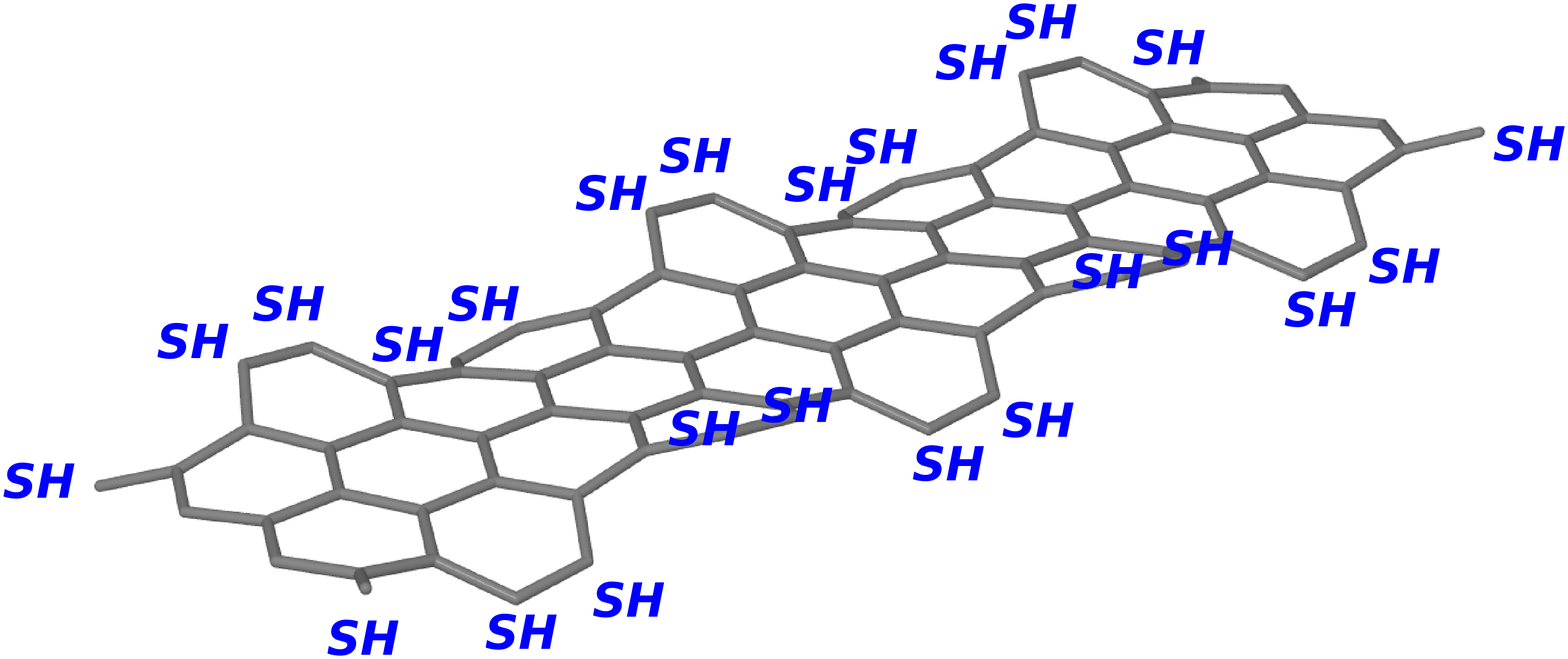}}
\subfigure [][] {\includegraphics[scale=.15]{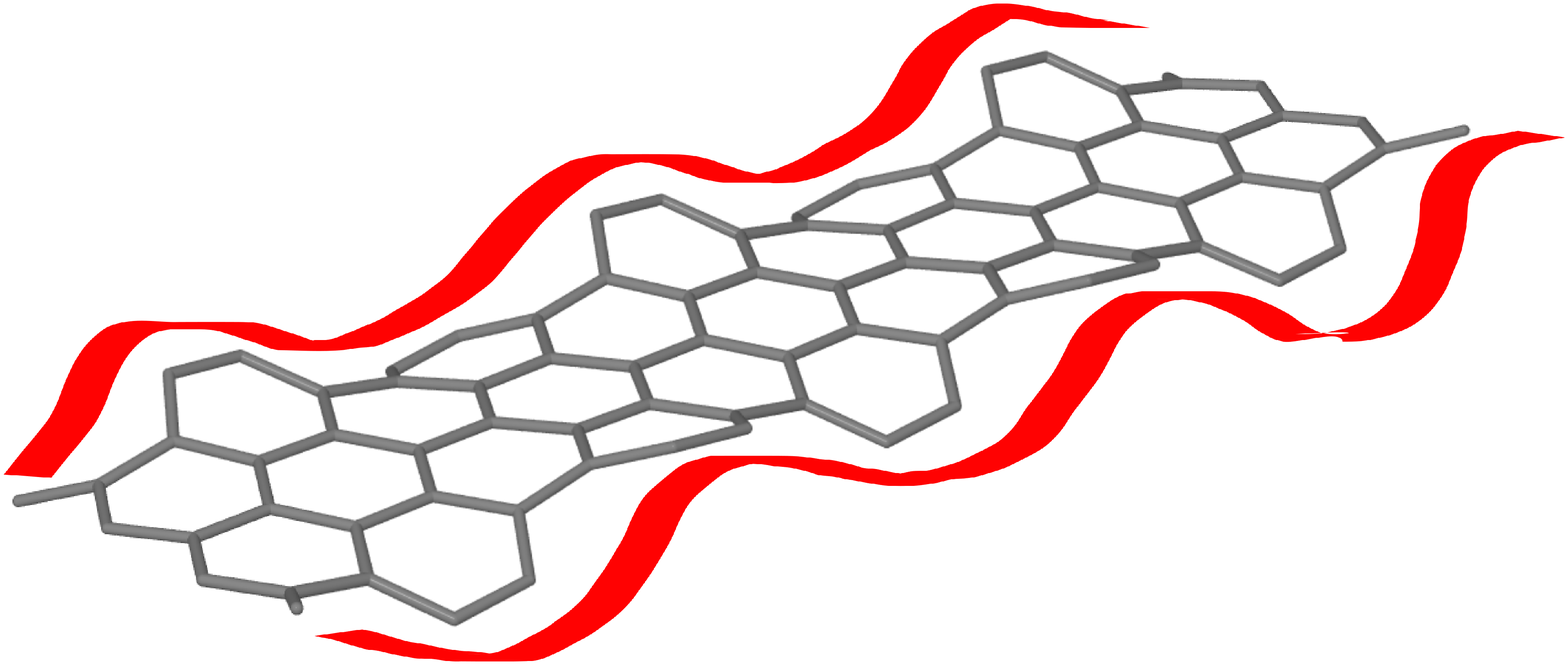}}

\caption{-SH terminated armchair graphene nanoribbon of width 7. (a) Rippled graphene ribbon edges with grafted SH-groups (in gray the graphene ribbon network, in blue the -SH groups, one -SH group per carbon edge atom), (b) as guide for the eye the effective rippled armchair graphene nanoribbon edges are shown by a thick red line.}
\label{figure_edge_ripple}       
\end{figure}

In Figure~\ref{figure_edge_ripple}a relaxed armchair graphene nanoribbon of width 7 with thiol functional groups is shown  (width nomenclature is taken from Reference \cite{Cervantes-Sodi2008}). A key consequence of these functionalised nanoribbon edges is that both electronic and mechanical properties can be tuned. By changing the functional groups the band gap can vary for certain ribbon widths up to 50 \%. Rippled structures can also decrease the Young's Modulus for small width graphene nanoribbons up to 40 \% compared to H-terminated armchair graphene nanoribbons \cite{Wagner2011}.\\

In general the influence of ``complex" functionalised graphene edges become increasingly important as the surface area to edge length ratio decreases \cite{Wagner2011}.  Thus edge rippling is likely to be of key importance in graphene nanoribbons with small widths ($< 2$ nm) and small diameter graphene flakes. 

\subsection{Twisting via functionalisation}

\begin{figure}[!]

\begin{center}
\includegraphics[scale=.20]{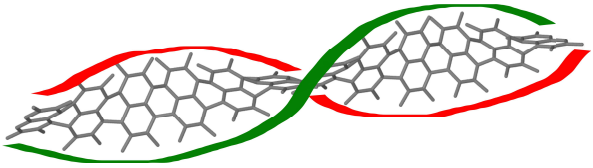}
\end{center}

\caption{Schematic armchair graphene nanoribbon width 7, twisted via $30\,^{\circ}$ between each original unit cell. The red and green line are guides for the eye indicating the two ribbon edges.  This twist is exaggerated, previous calculations find twist angles closer to 4$\,^{\circ}$ (see text)}
\label{twist_ripple}    
\end{figure}

An alternative way to relieve edge strain proposed recently is twisting of the whole graphene nanoribbon \cite{Gunlycke2010}. A schematic is shown in Figure~\ref{twist_ripple}. In this study a -F terminated armchair graphene nanoribbon of width 7 was twisted helically and found to be most stable with a twisting angle of $4.2 \,^{\circ}$ per unit cell. In this case the sp$^2$ $\pi$ - system of the graphene nanoribbon is  slightly distorted throughout compared to a flat graphene sheet. 

The literature study used tight binding and small cells with twisted boundary conditions, and it is not clear the cell size was sufficient \cite{Gunlycke2010}. We have therefore also calculated the structure and energy of such twisted ribbons using a large orthorhombic unit cell containing 1620 atoms (ribbon width 7) performing a $360^{\circ}$ twist in the unit cell with an angle of $4^{\circ}$ between fundamental unit cells, using a more accurate DFT-LDA method as implemented in the AIMPRO code \cite{Briddon2000}).  Such system sizes are possible due to development of a new filtration method for DFT calculations (for computational details see \cite{Rayson2009,Rayson2010}). First preliminary results suggest that twisted ribbons are less stable than the flat ribbons with rippled edges.  This seems sensible since edge rippling disturbs only the edge of the sp$^2$-graphene network whereas twisting distorts the entire ribbon.  Even if twisting can occur this suggests a limitation in width where twisting can be applied, as rippling the edge is independent of the graphene nanoribbon width.

\section{Conclusion}
We have discussed here a range of topological possibilities for graphene edges, many of which are unique to two-dimensional crystal lattices.  In-plane possibilities include rehybridisation, restucturing and reconstruction, all of which have been observed in HRTEM.  Out-of-plane distortions depend on the axis of distortion and include folding, scrolling, `tubing', as well as rippling and possibly even twisting once chemical functionalisation of the graphene edge sites is included.  Indeed it should also be possible to combine these orthogonal distortions, resulting in, for example, rippled edges which then form scrolls.  

We note that the discussion here applies to free-standing graphene.  Many topological distortions rely on a balance between an energetic cost to distort the sheet, offset by either strain relief or inter-layer interaction energy.  Adding a further surface energy via interaction with a substrate will alter these ratios, and at the very least change the zones of stability for each topological distortion.
The study of interfaces in two-dimensional materials opens up a rich diversity of structures which we have only begun to characterise and exploit experimentally.

\acknowledgements
This work has been carried out within the NANOSIM-GRAPHENE project n$^{\circ}$ANR-09-NANO-016-01 funded
by the French National Agency (ANR) in the frame of its 2009 programme in Nanosciences, Nanotechnologies and
Nanosystems (P3N2009).  We thank the COST Project MP0901 "NanoTP" for support.

\bibliographystyle{apsrev}
\bibliography{Edges_PHW}

\end{document}